\documentclass[sigconf,screen]{acmart}
\usepackage[nolist]{acronym}
\usepackage{tikz}
\usetikzlibrary{positioning,calc,backgrounds}
\usepackage{subcaption}
\usetikzlibrary{arrows.meta,positioning}
\usepackage{placeins}
\definecolor{darkgreen}{rgb}{0.0,0.45,0.0}
\definecolor{gold}{rgb}{1.0,0.84,0.0}
\definecolor{brown}{rgb}{0.55,0.27,0.07}
\definecolor{teal}{rgb}{0.0,0.5,0.6}
\definecolor{olive}{rgb}{0.45,0.45,0.0}
\definecolor{darkblue}{rgb}{0.0,0.0,0.55}
\definecolor{black}{rgb}{0.0,0.0,0.0}
\definecolor{salmon}{rgb}{0.98,0.5,0.45}
\usepackage{xurl} 
\usepackage[T1]{fontenc}
\usepackage{todonotes}

\newcommand{\api}[1]{\texttt{\hyphenchar\font=`\_ #1}}

\newcommand{\steward}[1]{\textcolor{black}{#1}}

\newcommand{\nathan}[1]{\textcolor{black}{#1}}

\newcommand{\discussionMarSix}[1]{\textcolor{black}{#1}}
\expandafter\newcommand\csname lawrence2\endcsname[1]{\textcolor{black}{#1}}
\newcommand{\meetingJuneNineteen}[1]{\textcolor{black}{#1}}
\newcommand{\reviewers}[1]{\textcolor{black}{#1}}

\newcommand{\intel}[1]{\textcolor{black}{#1}}

\newcommand{\ishmem}{Intel\textsuperscript{\textregistered} SHMEM}

\AtBeginDocument{%
  }

\makeatletter
\newcommand{\acg}[1]{%
	\expandafter\ifx\csname AC@\AC@prefix#1\endcsname\AC@used
		\acs{#1}'s%
	\else
		\aclu{#1}'s (\acs{#1}'s)%
	\fi
}

\newcommand{\acsg}[1]{\acs{#1}'s}

\makeatother

\begin{acronym}
\acro{AMD}{Advanced Micro Devices}
\acro{CPU}{central processing unit}
\acrodefplural{CPU}[CPUs]{central processing units}
\acro{GPU}{graphics processing unit}
\acrodefplural{GPU}[GPUs]{graphics processing units}
\acro{HPC}{high performance computing}
\acro{PGAS}{partitioned global address space}
\acro{RMA}{remote memory access}
\acrodefplural{RMA}[RMAs]{remote memory accesses}
\acro{SYCL}{single-source C++ heterogeneous programming}
\acro{GDA}{GPUDirect Async}
\acro{IPC}{inter-process communication}
\acro{Dedicated}{Dedicated communication resources}
\acro{Shared}{Shared communication resources}
\acro{xGMI}{external global memory interconnect}
\acro{PCIe}{peripheral component interconnect express}
\acro{QP}{Queue Pair}
\acrodefplural{QP}[QPs]{Queue Pairs}
\end{acronym}

\begin{document}

\acmConference[CUG '26]{The Cray User Group Annual Meeting}{April 26--30, 2026}{Nice, France}

\title{Toward a Unified GPU-Aware OpenSHMEM Specification}

\author{Naveen Ravi}
\orcid{0000-0003-3695-5696}
\email{nravi@hpe.com}
\author{Nathan Wichmann}
\email{wichmann@hpe.com}
\author{Md. Wasi-ur- Rahman}
\orcid{0000-0002-3465-4307}
\email{md.rahman@hpe.com}
\affiliation{%
  \institution{Hewlett Packard Enterprise (HPE)}
  \city{Bloomington}
  \state{Minnesota}
  \country{USA}
}

\author{Aurelien Bouteiller}
\email{Aurelien.Bouteiller@amd.com}
\author{Y{\i}ltan Hassan Temu\c{c}in} 
\orcid{0000-0002-4145-4848}
\email{yiltan@amd.com}
\author{Avinash Kethineedi}
\email{avinash.kethineedi@amd.com}
\author{Johnathan Alsop}
\email{Johnathan.Alsop@amd.com}
\author{Brandon Potter}
\email{Brandon.Potter@amd.com}
\affiliation{%
  \institution{Advanced Micro Devices, Inc.}
  \city{Austin}
  \state{Texas}
  \country{USA}
}

\author{Shubhendra Pal Singhal}
\orcid{0000-0002-0610-7672}
\email{ssinghal74@gatech.edu}
\author{Jun Shirako}
\orcid{0000-0002-7900-7680}
\email{shirako@gatech.edu}
\author{Akihiro Hayashi}
\orcid{0000-0001-6861-6272}
\email{ahayashi@gatech.edu}
\author{Vivek Sarkar}
\orcid{0000-0002-3433-8830}
\email{vsarkar@gatech.edu}
\affiliation{%
  \institution{Georgia Institute of Technology}
  \city{Atlanta}
  \state{Georgia}
  \country{USA}
}


\author{Lawrence C. Stewart}
\email{stewart@serissa.com}
\orcid{0000-0001-5149-1063}
\affiliation{%
  \institution{Serissa Research}
  \city{Wayland}
  \state{Massachusetts}
  \country{USA}
}

\author{Michael Beebe}
\email{michael.beebe@ttu.edu}
\orcid{0009-0004-9151-6607}
\affiliation{%
  \institution{Texas Tech University}
  \city{Lubbock}
  \state{Texas}
  \country{USA}
}

\author{Benjamin Michalowicz}
\email{michalowicz.2@osu.edu}
\orcid{0000-0002-7450-5787}
\affiliation{%
  \institution{The Ohio State University}
  \city{Columbus}
  \state{Ohio}
  \country{USA}
}

\author{Jeongnim Kim}
\email{jeongnim.kim@intel.com}
\orcid{0000-0002-0920-4515}
\author{Thiago Teixeria}
\email{thiago.teixeira@intel.com}
\orcid{0000-0002-8031-0652}
\author{Mark F. Brown}
\email{mark.f.brown@intel.com}
\orcid{0009-0009-2482-4497}
\affiliation{%
  \institution{Intel Corporation}
  \city{Hillsboro}
  \state{Oregon}
  \country{USA}
}


\author{Aaron Welch}
\email{welchda@ornl.gov}
\orcid{0000-0002-8988-3027}
\author{Oscar Hernandez}
\email{oscar@ornl.gov}
\orcid{0000-0002-5380-6951}
\affiliation{%
  \institution{Oak Ridge National Laboratory}
  \city{Oak Ridge}
  \state{Tennessee}
  \country{USA}
}

\author{Wendy Poole}
\email{wkpoole@lanl.gov}
\orcid{0009-0007-4984-6209}
\author{Steve Poole}
\email{swpoole@lanl.gov}
\orcid{0000-0002-4531-7453}
\affiliation{%
  \institution{Los Alamos National Laboratory}
  \city{Los Alamos}
  \state{New Mexico}
  \country{USA}
}

\renewcommand{\shortauthors}{Ravi et al.}

\begin{abstract}
Leadership-class HPC systems are now accelerator-centric, with GPUs providing most floating-point throughput and memory bandwidth.
As next-generation systems increasingly integrate accelerators through high-speed memory fabrics and system interconnects, exposing larger tightly coupled device domains, \ac{PGAS} models such as OpenSHMEM provide a natural abstraction for expressing fine-grained remote memory operations across these devices.
While OpenSHMEM 1.x offers a lean \acs{PGAS} model for irregular communication, atomics, fine-grained synchronization, and collectives, its memory model lacks portable semantics for accelerator architectures.
As a result, existing \acs{GPU}-enabled OpenSHMEM implementations differ in memory management, capability discovery, and operation semantics, limiting portability and ecosystem cohesion.
This risks fracturing the community that OpenSHMEM was originally created to unify.

This paper proposes an OpenSHMEM Auxiliary Specification for \acs{GPU}-Aware Communication, designed as a lightweight, backward-compatible extension to OpenSHMEM 1.x.
The auxiliary specification introduces a minimal memory model extension via a GPU-scoped memory space abstraction, along with capability queries and well-defined semantics for using \acs{GPU}-attached buffers in \acl{RMA}, atomic, synchronization, and collective operations.
This is initially conceived through the lens of a host-initiated interface, although it provides a general set of semantics that also allow for optional device-initiated support. A central goal of this effort is to demonstrate that GPU-aware OpenSHMEM semantics can be specified and implemented across GPUs from multiple vendors, providing a practical and rapidly implementable step toward unification under a vendor-neutral specification while informing the design of future OpenSHMEM specifications.

\end{abstract}





\maketitle

\section{Introduction}
\label{sec:intro}

Modern leadership-class \acs{HPC} systems rely on accelerators, with \acsp{GPU} contributing most of the computing capability and memory bandwidth.
For many workloads, performance and energy efficiency depend on keeping data resident in \acs{GPU} memory and minimizing data movement across memory boundaries.
At the same time, modern applications are growing in size and complexity, with increasing memory and communication demands across GPU devices.
Their communication patterns are diverse, including fine-grained remote updates in sparse linear algebra and graph analytics, structured halo exchanges in stencil-based solvers, FFT-driven spectral phases, many-to-many exchanges in data analytics and mixture-of-experts models, and latency-sensitive collectives in large-scale deep learning and AI workloads. These patterns frequently coexist within applications and must be composed efficiently with \acs{GPU} kernel execution.

The OpenSHMEM programming model has long been attractive for expressing such communication patterns.
As a \ac{PGAS} model, OpenSHMEM provides a small but expressive set of one-sided communication operations, atomic operations, synchronization primitives, and collectives operating on remotely accessible shared data objects.
Its design emphasizes explicit control, low software overhead, and predictable semantics, making it particularly well suited for irregular and fine-grained communication.
The OpenSHMEM 1.x specification formalized these semantics for \acs{CPU}-attached memory and has proven effective across a wide range of distributed-memory systems~\cite{openshmem16}.

However, heterogeneous architectures expose limitations in the OpenSHMEM memory model.
Today, remotely accessible memory is created during initialization, typically assigned a maximum size via an environment variable, and has implementation-defined characteristics.
On heterogeneous nodes, this raises fundamental questions: where should this memory reside, how many regions are needed, and how should memory placement be controlled?
More importantly, OpenSHMEM 1.x does not define portable semantics for using \acs{GPU}-attached memory as operands in communication and synchronization operations.
As \acsp{GPU} introduce distinct coherence and ordering domains, it becomes essential to define when data become visible to \acs{GPU} kernels, when remote updates are complete, and how communication interacts with device execution.

In response to these challenges, several \acs{GPU}-focused OpenSHMEM implementations have emerged.
NVIDIA’s NVSHMEM provides an OpenSHMEM-compatible interface for \acs{GPU}-resident communication, supporting host- and device-initiated operations on nodes with NVIDIA \acsp{GPU}~\cite{nvshmem}.
Similarly, \acsg{AMD} rocSHMEM enables \acs{GPU}-centric communication using OpenSHMEM-like semantics on \acs{AMD} \acs{GPU} platforms~\cite{rocshmem}. Intel SHMEM~\cite{brooks2024intel} brings similar capabilities to Intel \acs{GPU}s in a \ac{SYCL}-based environment~\cite{SYCL-2020-spec}.
These systems demonstrate both the demand for OpenSHMEM-style communication on \acsp{GPU} and the feasibility of achieving high performance with \acs{GPU}-resident \acp{RMA} operations.

Despite their success, existing \acs{GPU}-enabled OpenSHMEM implementations are not interchangeable.
They differ in memory management, how \acs{GPU} awareness is enabled, what capability discovery mechanisms exist, and what guarantees are provided when \acs{GPU}-attached buffers participate in \ac{RMA}, atomics, synchronization, and collectives.
These differences force vendor-specific code paths in applications, complicate portability, and make it difficult for performance tools and correctness analyzers to reason uniformly about program behavior. A key motivation for this auxiliary specification is to show that a single, well-defined GPU-aware OpenSHMEM interface can be specified and implemented across GPUs from multiple vendors.
We therefore frame the auxiliary specification to accommodate both CPU-first implementations (GPU-aware features enabled incrementally) and GPU-first implementations (GPU-resident communication as the primary path), so the baseline does not privilege one adoption model over the other.

This paper intentionally designs a narrowly scoped extension to the v1.x OpenSHMEM specification to enable \acs{GPU}-aware OpenSHMEM.
The objective is to provide a specification that can be adopted quickly, evaluated by vendors and users, and provides a low-risk learning vehicle.
\reviewers{The proposal is framed as a semantic contract rather than a replacement programming model: it defines memory placement, completion, ordering, and capability discovery rules that application teams, system software teams, and vendors can reason about consistently.}
A Tier-0 baseline should be implementable by at least two vendors, remain backward compatible with OpenSHMEM 1.x, and require conformance tests that can be integrated straightforwardly into existing OpenSHMEM test workflows.
Despite this intentionally limited scope, the technical challenges remain substantial.
In this paper we describe how to define semantics for memory visibility, preserve high performance in implementation, and establish a path for future extension.
Overall, this auxiliary specification is a crucial stepping stone toward a more ambitious unified \acs{GPU}-aware OpenSHMEM 2.0, whose scope and semantics can be informed by practical user experience and feedback gained here.

The goal of this work is to define an \emph{OpenSHMEM Auxiliary Specification for \acs{GPU}-Aware Communication}.
Rather than replacing OpenSHMEM 1.x or introducing a new programming model, the auxiliary specification extends OpenSHMEM 1.6~\cite{openshmem16} in a carefully scoped and backward-compatible manner.
All existing OpenSHMEM applications remain valid, while applications that require \acs{GPU}-resident communication gain a portable and well-defined interface.

The contributions of this paper are:
\begin{itemize}
  \item A clear taxonomy of GPU-aware OpenSHMEM features and a cross-implementation comparison grounded in evidence.
  \item A rationale for a vendor-neutral auxiliary specification that defines GPU-aware memory spaces, contexts, and operational semantics.
  \item A decision log and tiering strategy that balances portability with optional GPU-initiated capabilities.
  \item A conformance-oriented mapping from proposed specification to testable behaviors.
\end{itemize}

At the core of the proposal is a minimal extension to the OpenSHMEM memory model.
In addition to the remotely accessible memory provided by default, applications may collectively create at most one additional region, referred to as a \emph{memory space}.
A memory space is associated with an OpenSHMEM team and parameterized by explicit traits, including size and memory type (e.g., \acs{CPU}-attached or \acs{GPU}-attached memory).
Objects allocated from a memory space inherit its traits, enabling the explicit placement of data in \acs{GPU} memory while preserving OpenSHMEM’s allocation semantics.

Building on this foundation, the auxiliary specification defines \acs{GPU}-aware semantics for existing OpenSHMEM operations.
\Ac{RMA}, atomic operations, synchronization, and collectives are specified to accept \acs{GPU}-attached buffers under well-defined conditions.
Completion and ordering guarantees remain consistent with OpenSHMEM 1.6, while recognizing that applications must coordinate \acs{CPU} and \acs{GPU} execution using mechanisms provided by their \acs{GPU} programming environment.
This standardization enables portable reasoning about correctness and performance.

Capability discovery is another central design principle.
Because \acs{GPU}-aware communication support is not uniform across platforms, the auxiliary specification introduces explicit query interfaces that allow applications to determine support at runtime.
Environment-level controls further enable system-wide policy selection without application modification.
Together, these mechanisms support robust, adaptive applications across diverse systems.

Finally, the auxiliary specification defines a forward path toward \acs{GPU}-centric, kernel-initiated communication.
While device-side initiation is not mandated, the specification defines a common baseline, \acs{GPU}-attached memory and \acs{GPU}-aware semantics, that both host- and device-initiated implementations can build upon.
This incremental approach enables convergence today while leaving room for a more comprehensive OpenSHMEM 2.0 specification in the future.

For application developers, this enables portable GPU-resident communication. For implementers, it defines a low-risk convergence target. For tool developers, it establishes a stable semantic baseline.

\section{Background}
\label{sec:background}

This section introduces the OpenSHMEM programming model and the minimum GPU execution concepts needed to understand the auxiliary specification. The goal is to establish a common vocabulary without spending excessive space on tutorial material.

\subsection{OpenSHMEM basics: symmetric heaps and one-sided operations}
\label{sec:background:openshmem}

OpenSHMEM is a PGAS model where each processing element (PE) exposes a symmetric heap. A symmetric heap is a collectively allocated region that exists at each PE with the same size and allocation order, enabling remote memory access (RMA) with predictable addressing. OpenSHMEM provides one-sided put/get operations, atomic memory operations, collective operations, and synchronization primitives that operate on symmetric objects. The model emphasizes explicit control and low overhead by avoiding implicit messaging or rendezvous protocols in the application code path.

RMA semantics are defined in terms of completion, ordering, and visibility. A put updates remote memory, while a get fetches remote data into local memory. Atomic operations combine remote update with a compute operation (e.g., atomic add), and may return the value prior to the update (e.g., atomic fetch add). Synchronization calls (such as quiet, fence, wait/test, barrier, and team collectives) establish ordering and visibility relationships across PEs.

In the CPU-only OpenSHMEM model, these operations typically act on host-resident memory with a relatively strong and well-understood memory consistency model. Completion and ordering guarantees can often be reasoned about with host-side quiets and fences, and the runtime is assumed to make progress as long as the host thread continues to execute. GPU-aware OpenSHMEM must preserve these semantics while acknowledging that device memory and execution have different ordering and progress characteristics.

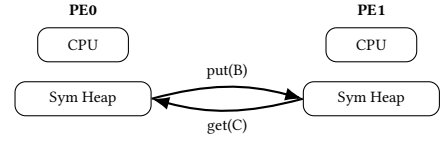
\begin{figure}[t]
  \centering
  \begin{tikzpicture}[>=Latex, font=\scriptsize]
    \tikzset{
      cpu/.style={draw, rounded corners, minimum height=0.45cm, minimum width=1.2cm, inner sep=2pt},
      heap/.style={draw, rounded corners, minimum height=0.45cm, minimum width=1.8cm, inner sep=2pt}
    }
    \node[cpu] (cpu0) {CPU};
    \node[heap, below=0.25cm of cpu0] (heap0) {Sym Heap};
    \node[cpu, right=2.6cm of cpu0] (cpu1) {CPU};
    \node[heap, below=0.25cm of cpu1] (heap1) {Sym Heap};
    \node[above=0.05cm of cpu0, font=\bfseries\scriptsize] {PE0};
    \node[above=0.05cm of cpu1, font=\bfseries\scriptsize] {PE1};

    \draw[-Latex, thick] (heap0.east) to[bend left=15] node[midway, above] {put(B)} (heap1.west);
    \draw[-Latex, thick] (heap1.west) to[bend left=15] node[midway, below] {get(C)} (heap0.east);
  \end{tikzpicture}
  \caption{Symmetric allocation enables direct remote addressing in OpenSHMEM.}
  \Description{Two processing elements, each with a CPU and symmetric heap. Arrows labeled put and get connect the heaps to show one-sided access.}
  \label{fig:symmetric-heap}
\end{figure}

Figure~\ref{fig:symmetric-heap} illustrates symmetric heaps and one-sided put/get operations between PEs.

\subsection{Teams, contexts, and completion/ordering}
\label{sec:background:contexts}
According to the OpenSHMEM 1.6 programming model~\cite{openshmem16} (Section 2), teams are subsets of PEs used for collective operations and for scoping point-to-point operations. Contexts provide an execution environment for communication operations, in which operations are ordered and completed independently from those issued in other contexts.
\steward{Teams also anchor atomicity domains through the contexts associated with them. The CPU specification already distinguishes predefined teams such as \api{SHMEM\_TEAM\_WORLD} and \api{SHMEM\_TEAM\_SHARED}. A GPU-centric extension therefore needs analogous device teams for world and locality scopes, so device-initiated and host-initiated atomics do not silently share an atomicity domain when hardware cannot provide that guarantee. Operations issued on different contexts remain unordered with respect to each other unless the specification states otherwise.}
This matters for GPUs because context handling determines what state is visible to device code and how operations are ordered across threads, blocks, or kernels. Many vendor-specific OpenSHMEM GPU implementations introduce explicit device contexts or thread-collaborative calls, making it important to define the expected scope of a context and its lifetime.

Completion and visibility are distinct concepts. Local completion means that the initiating PE can safely reuse the source buffer. Remote visibility means that the target PE can observe the update. Operations such as \texttt{fence} and \texttt{quiet} are used to enforce ordering and completion, and are critical when the underlying memory model is weak or when operations are deferred by the runtime. The auxiliary specification must therefore define how these concepts apply when data reside in GPU memory or when the issuer is a device thread.
\steward{On GPU-capable systems, completion also interacts with memory placement and cache visibility. GPU-resident symmetric memory must be remotely addressable and registered for OpenSHMEM operations. Managed, unified, or copy-staged buffers belong in the portable GPU symmetric heap only when the implementation exposes that behavior through traits or capability queries. Even for registered GPU memory, host or NIC writes may require explicit synchronization before a running or subsequent kernel can safely consume the data. Similarly, a device thread that issues a put relies on device-side ordering and a host-visible \texttt{quiet} or equivalent completion mechanism to make the operation complete from the host perspective. These cross-domain completion rules are a major source of divergence across implementations and a key focus of the auxiliary specification.}

Ordering across contexts and teams is another practical concern. For example, a host thread may issue puts on a context bound to \texttt{SHMEM\_TEAM\_WORLD} while a GPU kernel issues operations on a context bound to a sub-team. Without explicit ordering primitives, the relative visibility of these operations is not guaranteed. This is why the auxiliary specification treats contexts as first-class objects and requires clear documentation of their scope and ordering.

\begin{figure}[t]
  \centering
  \begin{minipage}{0.48\columnwidth}
    \centering
    \begin{tikzpicture}[>=Latex, font=\scriptsize, scale=0.84, transform shape]
      \tikzset{box/.style={draw, rounded corners, minimum height=0.45cm, minimum width=1.1cm, inner sep=2pt}}
      \node[box] (ctx) {CTX};
      \node[box, right=0.3cm of ctx] (qp) {Shared resources};
      \node[box, right=0.3cm of qp] (nic) {NIC};
      \draw[-Latex, thick] (ctx) -- (qp);
      \draw[-Latex, thick] (qp) -- (nic);
      \node[above=0.2cm of ctx, font=\bfseries\scriptsize] {\ac{Shared}-style};
    \end{tikzpicture}
  \end{minipage}
  \hfill
  \begin{minipage}{0.48\columnwidth}
    \centering
    \begin{tikzpicture}[>=Latex, font=\scriptsize, scale=0.84, transform shape]
      \tikzset{box/.style={draw, rounded corners, minimum height=0.45cm, minimum width=1.0cm, inner sep=2pt}}
      \node[box] (ctx2) {CTX};
      \node[box, right=0.3cm of ctx2] (qp0) {QP0};
      \node[box, below=0.2cm of qp0] (qp1) {QP1};
      \node[box, right=0.3cm of qp0] (nic2) {NIC};
      \draw[-Latex, thick] (ctx2) -- (qp0);
      \draw[-Latex, thick] (ctx2) -- (qp1);
      \draw[-Latex, thick] (qp0) -- (nic2);
      \draw[-Latex, thick] (qp1) -- (nic2);
      \node[above=0.2cm of ctx2, font=\bfseries\scriptsize] {\ac{Dedicated}-style};
    \end{tikzpicture}
  \end{minipage}
  \caption{\ac{Shared}-style contexts share resources, while \ac{Dedicated}-style contexts map to per-context queue pairs (QPs) or queues for isolation and tuning.}
  \Description{Side-by-side diagrams. \ac{Shared} has one context feeding a shared queue to a NIC. \ac{Dedicated} has one context feeding two queues to a NIC.}
  \label{fig:ipc-gda}
\end{figure}
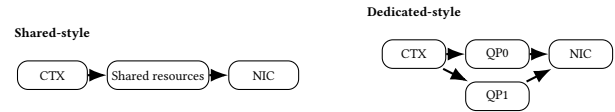

We use \ac{Shared}-style to denote shared-resource context models and \ac{Dedicated}-style to denote per-context resource models. Figure~\ref{fig:ipc-gda} contrasts \ac{Shared}-style contexts with \ac{Dedicated}-style contexts that allocate communication resources on a per-context basis, an optimization found in current GPU centric implementations.
In \ac{Shared}-style designs, such as the rocSHMEM \ac{IPC} backend, contexts do not explicitly map to fabric-level resources such as \ac{PCIe}~\cite{PCI-SIG-PCIe-Base} or \ac{xGMI} queues~\cite{AMD-xGMI}. Instead, \ac{Shared} contexts and teams use backend-managed shared \ac{IPC} pools for auxiliary state, including pSync, pWrk, and fence buffers. Transport endpoints and progress mechanisms are shared across contexts and managed by the backend. \intel{\ishmem{} is another \ac{Shared}-style implementation: context creation is implicit, no per-context endpoint resources are allocated, and GPU-initiated operations are issued on the default context with backend-managed shared resources.} In this model, each team or collective instance maintains its own synchronization and scratch state, but communication endpoints are not bound to individual contexts. In contrast, on InfiniBand, the rocSHMEM \ac{GDA} backend is an example of a \ac{Dedicated}-style backend that maps each context to a bank of \acp{QP}~\cite{InfiniBand-Architecture}, forming a fully connected topology in which each QP is assigned to a specific remote PE. This per-context allocation provides dedicated endpoints for GPU-initiated operations and enables stronger isolation and resource control at the cost of increased QP consumption. Similarly, NVSHMEM exposes explicit QP selection through the \texttt{nvshmemx\_qp\_create} and \texttt{nvshmemx\_qp\_*} APIs, allowing applications to explicitly choose which bank of QPs are used for a given operation. In short, \ac{Shared}-style designs share transport endpoints while maintaining per-team state, \ac{Dedicated}-style designs allocate per-context endpoint resources, and NVSHMEM exposes explicit mechanisms for endpoint steering~\cite{rocshmem,nvshmem}.

\subsection{GPU execution and memory spaces}
\label{sec:background:gpu}

GPUs execute many threads in single-instruction, multiple-thread (SIMT) fashion with a memory hierarchy that includes \intel{device, shared and local memory}. Device memory has weaker ordering guarantees than typical CPU memory, and kernel execution may be associated with streams or queues that define ordering among GPU operations. A GPU-aware OpenSHMEM model must specify how OpenSHMEM operations relate to stream ordering and device synchronization. When device threads initiate communication, explicit fences or quiet operations are often required to impose ordering on the weak device memory model.

We use two terms throughout the paper. (1) \texttt{GPU-aware communication} means that the runtime accepts GPU-resident buffers, but operations are initiated by the host. (2) \texttt{GPU-centric communication} means that GPU threads can initiate operations directly from kernels. Both approaches rely on symmetric memory and explicit ordering primitives, but GPU-centric support adds device-side APIs and stricter requirements on context visibility and progress. The auxiliary specification targets a baseline GPU-aware model while defining a path for optional GPU-centric capabilities.

GPU-capable platforms also introduce multiple memory spaces. Host memory can be pageable or pinned, device memory is attached to the GPU, and managed/unified memory provides a shared virtual address with runtime migration. Each space has different visibility and registration requirements for network access, and not all spaces are equally suitable for RMA or atomics or collectives. The auxiliary specification simplifies this landscape by defining a symmetric GPU-attached space with explicit traits and by requiring that GPU-resident operands be allocated from symmetric spaces, avoiding ambiguity about pointer validity or registration.

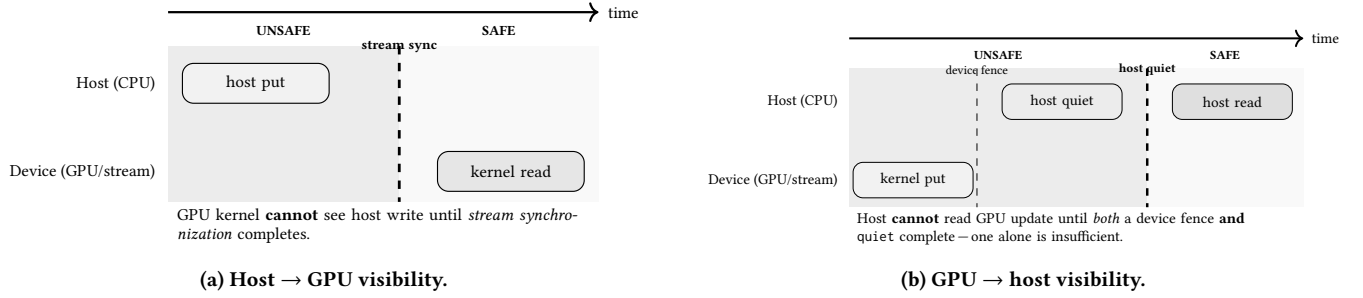
\begin{figure*}[t]
  \centering
  \begin{subfigure}[t]{0.48\textwidth}
    \centering
    \resizebox{\linewidth}{!}{%
    \begin{tikzpicture}[x=1.0cm,y=0.85cm,font=\scriptsize]
      \tikzset{
        lane/.style={draw=black!70, line width=0.8pt},
        event/.style={draw=black, fill=white, rounded corners, inner sep=2pt,
                      minimum height=5.0mm, text width=1.7cm, align=center},
        cpublock/.style={event, draw=black, fill=gray!10},
        safeblock/.style={event, draw=black, fill=gray!20},
        syncline/.style={black, dashed, line width=1.0pt},
        callout/.style={font=\scriptsize, align=left, text width=5.0cm, anchor=west},
      }
      \def\xstart{0.0}
      \def\xmax{5.4}
      \def\yhost{0.0}
      \def\ygpu{-1.3}
      \def\xsync{2.9}
      \def\yarrow{1.05}
      \def\yzone{0.55}

      \fill[gray!15] (\xstart,\ygpu-0.45) rectangle (\xsync,  \yzone);
      \fill[gray!5]  (\xsync,  \ygpu-0.45) rectangle (\xmax,   \yzone);

      \draw[->, thick] (\xstart,\yarrow) -- (\xmax,\yarrow) node[right]{\scriptsize time};

      \node[font=\tiny\bfseries]
        at ({(\xstart+\xsync)/2}, 0.78) {UNSAFE};
      \node[font=\tiny\bfseries]
        at ({(\xsync+\xmax)/2},   0.78) {SAFE};

      \node[anchor=east, font=\scriptsize] at (-0.05,\yhost) {Host (CPU)};
      \node[anchor=east, font=\scriptsize] at (-0.05,\ygpu)  {Device (GPU/stream)};

      \node[font=\tiny\bfseries, anchor=south]
        at (\xsync, 0.32) {stream sync};

      \draw[syncline] (\xsync,\yzone) -- (\xsync,\ygpu-0.45);

      \node[cpublock]  at (1.1, \yhost) {host put};
      \node[safeblock] at (4.3, \ygpu)  {kernel read};

      \begin{pgfonlayer}{background}
        \draw[lane] (\xstart,\yhost) -- (\xmax,\yhost);
        \draw[lane] (\xstart,\ygpu)  -- (\xmax,\ygpu);
      \end{pgfonlayer}

      \node[callout] at (\xstart,-2.10) {%
        GPU kernel \textbf{cannot} see host write until
        \textit{stream synchronization} completes.};
    \end{tikzpicture}%
    }
    \caption{Host $\rightarrow$ GPU visibility.}
  \end{subfigure}
  \hfill
  \begin{subfigure}[t]{0.48\textwidth}
    \centering
    \resizebox{\linewidth}{!}{%
    \begin{tikzpicture}[x=1.0cm,y=0.85cm,font=\scriptsize]
      \tikzset{
        lane/.style={draw=black!70, line width=0.8pt},
        event/.style={draw=black, fill=white, rounded corners, inner sep=2pt,
                      minimum height=5.0mm, text width=1.55cm, align=center},
        gpublock/.style={event, draw=black, fill=gray!10},
        syncblock/.style={event, draw=black, fill=gray!15},
        safeblock/.style={event, draw=black, fill=gray!25},
        syncline/.style={black, dashed, line width=1.0pt},
        synclinefaint/.style={black!60, dashed, line width=0.7pt},
        callout/.style={font=\scriptsize, align=left, text width=5.5cm, anchor=west},
      }

      \def\xstart{0.0}
      \def\xmax{6.4}
      \def\yhost{0.0}
      \def\ygpu{-1.3}
      \def\xfence{1.8}
      \def\xquiet{3.0}
      \def\xbarrier{4.2}
      \def\yarrow{1.05}
      \def\yzone{0.55}

      \fill[gray!15] (\xstart,\ygpu-0.45) rectangle (\xbarrier, \yzone);
      \fill[gray!5]  (\xbarrier,\ygpu-0.45) rectangle (\xmax,    \yzone);

      \draw[->, thick] (\xstart,\yarrow) -- (\xmax,\yarrow) node[right]{\scriptsize time};

      \node[font=\tiny\bfseries]
        at ({(\xstart+\xbarrier)/2}, 0.78) {UNSAFE};
      \node[font=\tiny\bfseries]
        at ({(\xbarrier+\xmax)/2},   0.78) {SAFE};

      \node[anchor=east, font=\scriptsize] at (-0.05,\yhost) {Host (CPU)};
      \node[anchor=east, font=\scriptsize] at (-0.05,\ygpu)  {Device (GPU/stream)};

      \node[font=\tiny, anchor=south]
        at (\xfence, 0.32) {device fence};
      \node[font=\tiny\bfseries, anchor=south]
        at (\xbarrier, 0.32) {host quiet};

      \draw[synclinefaint] (\xfence,   \yzone) -- (\xfence,   \ygpu-0.45);
      \draw[syncline]      (\xbarrier, \yzone) -- (\xbarrier, \ygpu-0.45);

      \node[syncblock] at (\xquiet, \yhost) {host quiet};
      \node[gpublock]  at (0.9,     \ygpu)  {kernel put};
      \node[safeblock] at (5.4,     \yhost) {host read};

      \begin{pgfonlayer}{background}
        \draw[lane] (\xstart,\yhost) -- (\xmax,\yhost);
        \draw[lane] (\xstart,\ygpu)  -- (\xmax,\ygpu);
      \end{pgfonlayer}

      \node[callout] at (\xstart,-2.10) {%
        Host \textbf{cannot} read GPU update until \textit{both}
        a device fence \textbf{and} \texttt{quiet}
        complete\,---\,one alone is insufficient.};
    \end{tikzpicture}%
    }
    \caption{GPU $\rightarrow$ host visibility.}
  \end{subfigure}

  \caption{Cross-domain ordering is not implicit. Visibility requires explicit
    host\kern0.1em$\leftrightarrow$\kern0.1emdevice synchronization.
    Shaded region indicates writes not yet visible. Dashed lines indicate
    synchronization barriers, and light region indicates safe reads.}

  \Description{Two side-by-side swimlane diagrams showing host–GPU memory ordering
    constraints. Synchronization barriers separate unsafe and safe regions
    where memory updates become visible.}

  \label{fig:ordering-timeline}
\end{figure*}

The differences from the CPU-only model are therefore threefold: (1) execution is asynchronous with respect to the host, (2) memory ordering is weaker and split across host and device domains, and (3) memory spaces have explicit placement and registration requirements. Establishing these distinctions in the background helps motivate why the auxiliary specification must define contexts, ordering, and space semantics more explicitly than OpenSHMEM 1.x. Figure~\ref{fig:ordering-timeline} illustrates the explicit host–device synchronization required for cross-domain visibility, with (a) showing host-to-device ordering and (b) showing device-to-host ordering.

\section{Related Work and Evolution of GPU-OpenSHMEM}
\label{sec:related}

GPU-aware OpenSHMEM has evolved through a sequence of prototypes and research efforts that explore how one-sided communication should map to accelerator memory and execution. This section summarizes those threads to motivate the auxiliary specification and highlight the path toward a vendor-neutral model.

Early work in the 2015--2018 timeframe introduced GPU-aware symmetric memory allocation and Compute Unified Device Architecture (CUDA)-aware data paths using GPUDirect remote direct memory access (RDMA). These efforts demonstrated that one-sided semantics can be preserved while reducing host staging overhead, but they also exposed gaps in the OpenSHMEM 1.x specification around device memory, ordering, and completion. Subsequent studies noted that GPU memory models are weakly ordered, motivating explicit ordering primitives when applying OpenSHMEM semantics on accelerators. The lessons from these prototypes shaped later vendor implementations, especially around explicit ordering and memory registration requirements.

By the late 2010s, the community also explored how to express GPU-resident symmetric heaps and how much of OpenSHMEM could remain unchanged while supporting GPU pointers. This period clarified the need to separate a portable semantic baseline (symmetry, completion, ordering) from performance-specific implementation details such as registration caches and GPU-direct access paths. These results fed directly into current implementation strategies and the rationale for an auxiliary specification rather than an immediate full standard revision.

The first widely used GPU-aware OpenSHMEM implementation was NVSHMEM, which operationalized GPU-resident symmetric memory and device-initiated operations on NVIDIA systems~\cite{nvshmem}. Its deployment established that OpenSHMEM-style semantics can be preserved on GPU-resident data while still enabling high-performance GPU-initiated communication.

Subsequent research from 2018--2022 explored GPU-initiated communication more broadly, where device threads issue RMA and synchronization operations directly. This work demonstrated strong performance benefits from overlapping computation and communication, but it also required careful handling of device memory ordering and progress. rocSHMEM extends OpenSHMEM for AMD GPUs with device-side APIs and cooperative variants for wavefronts and workgroups, illustrating how SIMT execution influences API design and collective requirements~\cite{rocshmem}. The requirement that all threads in a wavefront or workgroup participate uniformly in collective device calls is a key constraint that the auxiliary specification must acknowledge.

\steward{Recent work on GPU-initiated communication and NVSHMEM applications has also motivated persistent or long-lived kernel patterns, where GPU threads maintain communication roles across iterations~\cite{hamidouche_ppopp2020,nvshmem}. These patterns amplify the need for clear progress guarantees and well-defined completion semantics because device-side operations may execute for long durations without host intervention. They also reveal that a portable API must clearly document restrictions on device participation and synchronization in order to avoid deadlocks in real applications.}

\steward{Intel SHMEM brings the same GPU-aware and GPU-centric questions into a \acs{SYCL} programming environment. It supports host- and device-initiated operations, but the context model and backend integration differ from the CUDA and HIP implementations. Taken together, NVSHMEM, rocSHMEM, and Intel SHMEM demonstrate both the feasibility and the variability of GPU-aware OpenSHMEM implementations.}

More recent work has argued for vendor-neutral abstractions and \acs{SYCL}-based portability layers. These approaches could unify cooperative GPU semantics across vendors. They reinforce the need for an auxiliary specification that defines baseline semantics while accommodating optional device-initiated capabilities.

\acs{SYCL} and similar portability layers are valuable for single-source programming, but they do not define PGAS-level semantics such as symmetric heaps, remote completion, or ordered one-sided operations across processes. As a result, they cannot substitute for a GPU-aware OpenSHMEM specification. Instead, they motivate a clear semantic contract that such frameworks can target.

Since 2022, community discussions have increasingly emphasized portability and tool accessibility: applications and performance tools need a stable semantic baseline to reason about correctness across vendors, while allowing optional extensions for advanced hardware features. The auxiliary specification proposed here is intended to capture this balance by defining portable GPU-aware behavior and documenting where implementations may provide additional device-centric features.

\subsection{MPI GPU-awareness alternatives}
We do not base our design on MPI because portable GPU support remains both implementation- and transport-dependent. \intel{Several MPI stacks accept CUDA/HIP/SYCL device pointers, but coverage for GPU-resident one-sided communication is inconsistent. Open MPI's CUDA-aware UCX path excludes RMA and window creation~\cite{mpi41,openmpi_cuda_ucx_limits}. MVAPICH2-GDR advertises MPI-3 RMA/atomics over GPU buffers~\cite{mvapich2gdr_features}. Intel MPI~\cite{intel_mpi_gpu_support} and MPICH~\cite{mpich_exa} support SYCL device pointers for point-to-point and collectives on Intel GPU platforms, and deliver GPU RMA over OFI/libfabric. Moreover, GPU communication with MPI variants currently remains host-orchestrated, inducing kernel-boundary synchronization and limiting strong scaling in GPU-resident phases. Device-initiated operations in SHMEM implementations target fine-grained in-kernel communication and have demonstrated speedups over GPU-aware MPI in stencil and graph workloads~\cite{nvshmem,potluri_hipc17_gpuverbs,potluri_bfs_nvshmem}.}
\steward{Specialized GPU collective libraries also target accelerator-resident communication. They are complementary to the auxiliary specification because they optimize collective communication patterns, while OpenSHMEM must also define PGAS-level symmetric heaps, RMA, atomics, synchronization, and completion semantics.}

\section{Implementation Challenges and Divergence}
\label{sec:divergence}

This section summarizes where current GPU-enabled OpenSHMEM implementations diverge and why those differences matter for portability.

\subsection{Context handling models}
\label{sec:divergence:contexts}

Current implementations differ in how they expose contexts to host and device code. Some provide implicit default contexts that are available to device kernels, while others also offer explicit context creation with device-visible handles.
\steward{Granularity varies primarily in the device API surface rather than in the context object itself.}
\csname lawrence2\endcsname{Thread participation within a PE is a separate concept from context granularity. A context defines communication state, ordering, resources, and lifetime. A participation scope defines which GPU threads collaborate to issue a single operation on behalf of the PE. For RMA, threads may issue independent operations or collaborate on one block of data. For collectives, the operation remains a PE collective ordered on a team, while GPU threads may collaborate only within the stated thread scope of a single collective invocation.}

A related divergence is whether contexts map to uniform shared resources or to per-context resources such as queue pairs or work queues. The latter enables isolation and tuning but increases setup cost and resource pressure. The auxiliary specification needs to abstract these differences into a portable model without hindering high-performance implementations.

For example, NVSHMEM exposes queue-pair (QP) APIs to map traffic to specific communication resources but does not provide public explicit context handles, while rocSHMEM exposes explicit context creation and emphasizes workgroup- and wavefront-scoped device APIs that imply a different participation scope for device operations~\cite{nvshmem,rocshmem}. \intel{In contrast, \ishmem{} exposes no explicit context creation or device-visible context handles for device-initiated communication, leaving context specialization unavailable at the device level in current releases, though this remains an active area of development~\cite{brooks2024intel}.} These divergences motivate a baseline model that permits explicit contexts but does not require per-context resource specialization.

\steward{Contexts, implicit or explicit, may also be applied to direct GPU--GPU fabrics such as NVLink, Infinity Fabric, and Xe Link. These load-store fabrics provide direct hardware-assisted communication among GPUs and may scale to the node or rack level. They can offer high-performance remote atomic operations, but not necessarily in the same atomicity domain as host load-store or NIC-assisted atomics. The OpenSHMEM specification defines atomicity domains through teams and their default contexts. A GPU-aware auxiliary specification therefore needs device teams such as \texttt{SHMEM\_DEVICE\_TEAM\_WORLD} and \texttt{SHMEM\_DEVICE\_TEAM\_SHARED}. These teams manage device atomicity domains while preserving implementation freedom for high-performance GPU atomics. GPU vector or block atomics are useful future extensions, but they are outside the baseline scope of this paper.}

Another divergence is how contexts interact with teams and collectives.\intel{NVSHMEM, rocSHMEM, and \ishmem{} all provide device-initiated collectives. NVSHMEM uses explicit \texttt{\_warp} and \texttt{\_block} variants. rocSHMEM uses explicit \texttt{\_wg} and \texttt{\_wave} variants. \ishmem{} uses \texttt{\_work\_group} variants. \csname lawrence2\endcsname{For \ishmem{}, the explicit \texttt{sycl::group} argument exposes the relevant SYCL group object to lower-level implementation routines. It should not be read as a distinct context mechanism or as the source of the participation rule.}\footnote{\ishmem{}'s \texttt{\_work\_group} collective API requires a \texttt{sycl::group} argument, covering both workgroup and subgroup scopes.}} \steward{These are PE collectives invoked from device code. The GPU-specific question is thread collaboration within a PE: whether one thread, a warp or wave, a subgroup, a workgroup, or an entire grid must participate uniformly in each call. The auxiliary specification should define minimum participation rules for device collectives and leave richer thread-collaborative variants as optional capabilities.}

\subsection{Memory spaces and registration}
\label{sec:divergence:spaces}

All implementations rely on symmetric allocation, but the location and lifecycle of symmetric memory differ. Some systems allocate only a default symmetric heap, while others allow an additional GPU-targeted space. Memory registration requirements also differ across transports, affecting whether the runtime can expose GPU memory directly to the NIC or must stage through host buffers. These differences surface in the programming model as constraints on which pointers are valid for RMA and collectives.

\steward{NVSHMEM device operations address remote operands through symmetric-heap offsets, and its device collectives operate on buffers passed to team collective routines~\cite{nvshmem}. Some NVSHMEM collective paths first pack or copy the root PE's local source into team scratch before the inter-PE transfer, so this implementation detail should not be read as a general requirement that every local source buffer is itself remotely addressable. This does not imply that every local source buffer for device RMA must be symmetric. The portable rule in the current auxiliary draft is stricter for GPU-centric operations: source and destination operands reside in the GPU symmetric heap unless an implementation documents an extension. rocSHMEM follows symmetric allocation rules for its device API, and \ishmem{} exposes symmetric allocation for device-initiated operations~\cite{rocshmem,brooks2024intel}. These constraints suggest that the auxiliary specification should explicitly define which operands must be symmetric, rather than relying on implementation heuristics.}

Memory registration and direct access also differ across transports. NVSHMEM exposes \texttt{nvshmem\_ptr} to indicate when a remote object is directly addressable (e.g., via NVLink), \intel{rocSHMEM provides \texttt{rocshmem\_ptr} on both host and device, and \ishmem{} provides \texttt{ishmem\_ptr} similarly.} \csname lawrence2\endcsname{These vendor-prefixed routines are functional analogs of \texttt{shmem\_ptr}. Their names reflect each implementation's namespace rather than different semantic categories. A standard auxiliary interface should therefore define the common direct-pointer semantics instead of treating the prefixes as distinct concepts.} \intel{All three return a remote symmetric heap device pointer (or \texttt{NULL} if unavailable) to enable direct load/store from custom kernels~\cite{rocshmem,brooks2024intel}.}
Implementation-specific details are hidden behind the API but influence performance and availability of direct loads/stores.

\begin{table*}[tb]
  \centering
  \caption{Comparison of host-side stream/queue ordering semantics across GPU-enabled OpenSHMEM implementations}
  \label{tab:stream-ordering}
  \scriptsize
  \setlength{\tabcolsep}{3pt}
  \begin{tabular}{@{}lcccccc@{}}
    \toprule
    Impl. &
    RMA &
    Collectives &
    Sync &
    Completion &
    Abstraction &
    Device ordering \\
    \midrule
    NVSHMEM  &
    Yes &
    Yes &
    Yes (on-stream APIs) &
    Full &
    CUDA stream &
    No \\
    
    rocSHMEM &
    Yes &
    Partial &
    Partial &
    Partial (no quiet) &
    HIP stream &
    No \\
    
    \ishmem{}   &
    Yes &
    Yes &
    Yes (on-queue APIs) &
    Full &
    SYCL queue &
    No \\
    \bottomrule
  \end{tabular}
  \label{table:stream}
\end{table*}

\subsection{Completion, ordering, and progress}
\label{sec:divergence:ordering}

GPU memory models are weakly ordered, and device-initiated operations often relax ordering unless the programmer inserts explicit fences. Implementations generally expose \texttt{fence} and \texttt{quiet} to restore ordering and completion, but they differ in whether these operations imply remote visibility or only local completion. Stream-ordered variants further complicate behavior: GPU kernels may enqueue operations in different streams, and mixing collectives or waits across streams can introduce deadlock patterns if ordering is not carefully managed.

Progress also varies. Some runtimes rely on host proxies or NIC offload. Others allow GPU-driven progress on supported hardware. These differences affect persistent kernels and whether completion requires host participation.

NVSHMEM, rocSHMEM, and \ishmem{} all expose device fences and quiet operations, but their effective ordering semantics reflect the underlying GPU memory model and transport capabilities. The auxiliary specification therefore needs to separate ordering guarantees from performance optimizations, while still allowing implementations to document any required synchronization patterns (e.g., host/device stream interaction constraints).

Stream interaction is another source of divergence. Table~\ref{tab:stream-ordering} shows the comparison of host-side stream/queue ordering semantics across vendor-specific GPU OpenSHMEM implementations. NVSHMEM provides stream-ordered variants and documents hazards where collectives or waits can deadlock if streams are ordered inconsistently. \intel{\ishmem{} exposes host-side SYCL queue-ordered \texttt{\_on\_queue} APIs, where the choice of in-order or out-of-order queue and dependency handling are left to the user. Both NVSHMEM and rocSHMEM provide host-side \texttt{\_on\_stream} variants for put/get, barrier, broadcast, alltoall, and signal wait/put. \ishmem{} provides equivalent \texttt{\_on\_queue} variants. All implementations enqueue operations on a stream or queue and require synchronization for completion~\cite{rocshmem,nvshmem,brooks2024intel}.} \steward{None of the implementations exposes a device API that takes a stream or queue parameter. The auxiliary specification should therefore treat stream ordering as a host-issued feature and require that implementations document prohibited patterns.}. 


\subsection{Concrete examples}
\label{sec:divergence:examples}

Case study 1: GPU-resident halo exchange in a multi-GPU stencil code. One implementation allows device-initiated puts directly from GPU memory, but requires symmetric allocations and explicit stream ordering to avoid races between the kernel and the communication stream. Another implementation supports device-initiated variants via a host proxy rather than native GPU-initiated transport, which changes overlap potential. The auxiliary specification needs to make the symmetric allocation requirement explicit while clarifying how stream-ordered operations relate to completion and visibility for GPU kernels~\cite{nvshmem}.

Case study 2: fine-grained updates in graph analytics. GPU kernels may issue atomics or small puts to remote adjacency data while other threads poll completion flags. Some implementations expose device-side atomics but differ in ordering guarantees or allowable target locations, leading to portability hazards when a kernel assumes that atomics imply ordering for subsequent reads. A portable specification should separate atomicity from ordering and require explicit fence or quiet operations when ordering is needed, even if some implementations provide stronger behavior by default.

Case study 3: thread-collaborative collective reductions. \intel{NVSHMEM, rocSHMEM, and \ishmem{} all provide device-callable collective operations. rocSHMEM exposes \texttt{\_wg} and \texttt{\_wave} variants, NVSHMEM exposes block-level variants, and \ishmem{} exposes \texttt{\_work\_group} variants for SYCL groups~\cite{rocshmem,nvshmem,brooks2024intel}.} \steward{Applications that embed such collectives inside cooperative kernels must structure control flow so that the required participating threads call the operation uniformly. The auxiliary specification should define the minimum participation rules for PE collectives invoked from device code and allow richer thread-collaborative variants as optional capabilities.}

\section{Feature Taxonomy and Cross-Implementation Comparison}
\label{sec:comparison}

This section distills the comparison into a compact taxonomy and two tables: a feature matrix (Table~\ref{tab:taxonomy}) and a context-model comparison (Table~\ref{tab:context-model}). \nathan{The feature matrix separates the umbrella GPU-aware/centric model from three execution paths: host-executed, stream-triggered, and device-initiated.} \nathan{Stream-triggered APIs are still host-issued. They add ordering with a GPU stream or queue, but they do not make the GPU thread the issuing agent.} The context-model table highlights implicit/explicit contexts and whether resources are shared or per-context. Together, these tables show where a vendor-neutral baseline is feasible and where staged optional capabilities are needed.

The taxonomy is intentionally small: contexts, memory spaces, RMA operations, completion/ordering, atomics, synchronization, teams/collectives, progress, streams, and error handling. Table~\ref{tab:taxonomy} provides the abbreviated taxonomy used throughout the paper. We use AMO to denote atomic memory operations in the taxonomy and table headings.
\begin{table}[tb]
  \centering
  \caption{Feature taxonomy.}
  \label{tab:taxonomy}
  \footnotesize
  \setlength{\tabcolsep}{3pt}
  \begin{tabular}{@{}p{0.30\columnwidth}p{0.60\columnwidth}@{}}
    \toprule
    Category & Scope \\
    \midrule
    Contexts & Implicit/explicit, granularity, Shared vs Dedicated \\
    Spaces & CPU/GPU symmetric heaps and traits \\
    RMA/AMO & Put/get, atomics, ordering/completion \\
    Sync/Teams & Wait/test/signal, collectives, teams \\
    Progress/Streams & Progress model, stream ordering \\
    \bottomrule
  \end{tabular}
\end{table}
Table~\ref{tab:feature-matrix} expands the taxonomy into a compact support matrix, highlighting a baseline column that shows explicitly when a particular capability is to be optional in the auxiliary specification.
\nathan{Here, \emph{host-executed} means host code directly invokes the operation, \emph{stream-triggered} means host code enqueues it into a GPU stream or queue possibly with dependencies on kernel completions, and \emph{device-initiated} means the GPU thread issues it directly.} \nathan{All three paths are GPU-aware, but only the device-initiated path is GPU-centric in the stronger sense used later in the auxiliary specification.}
We use \emph{context specialization} to mean any API-visible mechanism that allows applications to steer operations to distinct communication resources or ordering domains, including explicit context handles (rocSHMEM) and explicit QP selection (NVSHMEM via \api{nvshmemx\_qp\_create} and \api{nvshmemx\_qp\_*}) even when a public context object is not exposed~\cite{rocshmem,nvshmem}.
\begin{table}[tb]
  \centering
  \caption{Implementation feature matrix}
  \label{tab:feature-matrix}
  \footnotesize
  \setlength{\tabcolsep}{3pt}
  \begin{tabular}{@{}p{0.32\columnwidth}cccc@{}}
    \toprule
    Feature & \ishmem{} & NVSHMEM & rocSHMEM & Baseline? \\
    \midrule
    GPU-aware buffers & Yes & Yes & Yes & Yes \\
    Device-initiated ops & Yes & Yes & Yes & Optional \\
    Stream/queue ordering & Yes & Yes & Partial & Optional \\
    Atomics on GPU targets & Yes & Yes & Yes & Yes \\
    Context specialization & Partial & Yes & Yes & Optional \\
    \bottomrule
  \end{tabular}
\end{table}
Table~\ref{tab:context-model} narrows the comparison to context semantics, including implicit/explicit availability, device-visible handles, and whether per-context resources are exposed.
In this table, \emph{per-context resources} means that a distinct transport endpoint set is allocated per context (e.g., a per-context QP bank). We label NVSHMEM as ``QP-only'' because it exposes QP selection without a public context handle. 
Implementations without per-context transport endpoint exposure are marked ``No''~\cite{rocshmem,nvshmem}.
\footnote{In \ishmem{}, the public header defines only a team config field (\texttt{num\_contexts}) and no explicit context create/destroy API. The team-split path stores this config without allocating per-context transport endpoints, so per-context resources are marked ``No'' (see \texttt{ishmem/src/ishmem.h} and \texttt{ishmem/src/teams.cpp}).}
\begin{table}[tb]
  \centering
  \caption{Context model comparison}
  \label{tab:context-model}
  \footnotesize
  \setlength{\tabcolsep}{3pt}
  \begin{tabular}{@{}lccc@{}}
    \toprule
    Aspect & Intel SHMEM & NVSHMEM & rocSHMEM \\
    \midrule
    Implicit context & Yes & Yes & Yes \\
    Explicit contexts & No & No & Yes \\
    Device-visible handle & No & No & Yes \\
    Per-context resources & No & QP-only & Yes \\
    \bottomrule
  \end{tabular}
\end{table}
Table~\ref{tab:api-summary} contrasts the three execution paths just defined, which anchors the separation between general GPU-aware capabilities and the GPU-centric device-initiated path used in the proposal section.
\begin{table}[tb]
  \centering
  \caption{API summary by execution path}
  \label{tab:api-summary}
  \scriptsize
  \setlength{\tabcolsep}{1pt}
  \begin{tabular}{@{}p{0.18\columnwidth}p{0.24\columnwidth}p{0.24\columnwidth}p{0.22\columnwidth}@{}}
    \toprule
    Category & Host-executed & Stream-triggered & Device-initiated \\
    \midrule
    Allocation & sym. heap/space & sym. heap/space & sym. heap/space \\
    RMA put/get & host call & stream call (opt.) & device call (opt.) \\
    Atomics & host call & stream call (opt.) & device call (opt.) \\
    Collectives & host collective & stream coll. (opt.) & device coll. (opt.) \\
    Ordering & fence/quiet & stream + fence/quiet & device fence/quiet \\
    \bottomrule
  \end{tabular}
\end{table}

\section{Proposed GPU-OpenSHMEM Auxiliary Specification}
\label{sec:proposal}

This section presents the proposed GPU-aware semantics and application programming interfaces (APIs) as a portable, backward-compatible extension to OpenSHMEM 1.x. Normative requirements are highlighted in callout excerpts and summarized in Table~\ref{tab:baseline-reqs}. Surrounding text provides rationale and design intent. The intent is to provide a clear rationale narrative rather than a full formal specification. \reviewers{Detailed routine signatures and illustrative code examples are maintained in the auxiliary specification draft~\cite{gpu_aux_spec_draft}. Here, we emphasize the portable semantic contract and the decision points behind it.}

\begin{table}[t]
  \centering
  \caption{Baseline requirements summary}
  \label{tab:baseline-reqs}
  \begin{tabular}{lp{0.58\columnwidth}}
    \toprule
    Rule & Requirement \\
    \midrule
    Symmetric GPU space & GPU-resident operands MUST be allocated from a symmetric space bound to the target team. \\
    Ordering primitives & \texttt{fence}/\texttt{quiet} MUST restore OpenSHMEM ordering/completion semantics for GPU-resident operations. \\
    Context visibility & Device-visible contexts MUST preserve host ordering and lifetime semantics. \\
    Device collectives & If provided, device collectives MUST require uniform participation in the stated scope. \\
    Capability queries & Optional features SHOULD be discoverable via capability queries. \\
    \bottomrule
  \end{tabular}
\end{table}

\subsection{Symmetric Memory Heap model}
\label{sec:proposal:spaces}

The auxiliary specification introduces a minimal two-space memory model: the default symmetric heap and at most one additional memory heap on GPU.
This additional heap is optional (that is, implementations may or may not provide it), and its creation can be controlled by environment variables specifying its critical traits.
User programs can obtain symmetric memory from this GPU heap by calling new APIs like \texttt{shmemg\_malloc} that are direct counterparts to the familiar operations on the CPU heap.
This design keeps the API small while enabling explicit placement of GPU-resident data. Managed or unified memory is outside the baseline model unless explicitly supported via traits or capability queries. Optional traits MAY include coherency or consistency hints (e.g., coarse/fine-grain or uncached), but the baseline only requires size and memory type. Any additional traits must be discoverable via capability queries. Heap granularity remains an implementation performance knob rather than part of the minimum portable contract. A key requirement is that allocation order and size remain symmetric across PEs for each space, preserving predictable, computable address offsets and interoperability with existing OpenSHMEM semantics.
GPU-first implementations may choose to create a GPU-resident symmetric heap at initialization.
When a GPU heap is available, new query APIs expose that capability with a new API \texttt{shmem\_query\_capabilities} that generalizes the concept of querying beyond just thread safety support.
The auxiliary specification assumes static heaps 
and a two-heap model (CPU + GPU). It realizes the portable GPU space as a single optional GPU symmetric heap.
\discussionMarSix{Implementations may also expose initialization-time controls for selecting feature subsets or default-heap behavior, but the portability contract relies on capability queries rather than a single required initialization API shape.}
Figure~\ref{fig:spaces} summarizes the baseline two-space model and contrasts it to the team-scoped spaces allocation rules that have been deferred to OpenSHMEM 2.0.

The restriction to a single additional space is a baseline requirement rather than a claim that multiple spaces are undesirable. One extra GPU heap is sufficient to cover the dominant GPU-aware use cases (host-attached and GPU-attached symmetric heaps) while avoiding the complexity of per-space registration keys, pointer tagging, and ordering rules that would otherwise multiply across transports. \steward{The baseline assumes a well-defined PE-to-GPU association. Implementations with multiple GPUs per PE or PEs without GPU association must expose the active association and device-team membership through capability queries. A model that treats GPUs as independent PEs is deferred to OpenSHMEM 2.0.}

By limiting the number of heaps, the specification avoids dynamic space proliferation and reduces complexity for registration and key management in transport backends. A single additional heap also avoids per-allocation overheads for identifying which heap a pointer belongs to (e.g., heap-base disambiguation or pointer-attribute queries) when multiple heaps exist. The baseline heap model therefore supports portability while leaving room for advanced implementations that can specialize registration or provide additional allocation hints.

As a possible future extension (deferred to OpenSHMEM 2.0), spaces may be created collectively by a team to preserve symmetry and avoid per-PE memory regions. Objects could inherit the traits of their space, enabling predictable reasoning about GPU-aware operations. Such spaces would likely be created infrequently and have immutable traits. Optional creation or destruction could occur at quiescent points, with allocation failures treated as collective errors.
\begin{figure}[t]
  \centering
  \fbox{\begin{minipage}{0.9\columnwidth}
  \small
  \textbf{Two-space model (baseline).}\\
  \textbf{CPU space:} default symmetric heap (host-attached)\\
  \textbf{GPU space:} optional symmetric heap with GPU-attached traits\\
  \textbf{Mspace model (OpenSHMEM 2.0).}\\
  \textbf{Team binding:} spaces are created collectively per team
  \end{minipage}}
  \caption{Baseline memory spaces model for GPU-aware OpenSHMEM.}
  \Description{A text callout describing the baseline two-space model with a default CPU space and an optional GPU space created per team.}
  \label{fig:spaces}
\end{figure}

\begin{figure*}[t]
  \centering
  \begin{subfigure}[t]{0.48\textwidth}
    \centering
    \resizebox{\linewidth}{!}{%
      \begin{tikzpicture}[font=\scriptsize, node distance=12mm, >=Latex]
        \tikzset{
          flow/.style={draw, rounded corners, minimum height=6mm,
                       text width=2.6cm, align=center},
        }
        \node[flow] (init) {shmem\_init()};
        \node[flow, right=of init] (implicit) {Create implicit ctx};
        \node[flow, right=of implicit] (publish) {Publish ctx handle\\(GPU-visible)};
        \draw[->, shorten >=2pt, shorten <=2pt] (init) -- (implicit);
        \draw[->, dashed, shorten >=2pt, shorten <=2pt]
          (implicit) -- node[midway, above, sloped]{publish} (publish);
      \end{tikzpicture}%
    }
    \caption{Host-side context creation and publication}
  \end{subfigure}
  \hfill
  \begin{subfigure}[t]{0.48\textwidth}
    \centering
    \resizebox{\linewidth}{!}{%
      \begin{tikzpicture}[font=\scriptsize, node distance=12mm, >=Latex]
        \tikzset{
          flow/.style={draw, rounded corners, minimum height=6mm,
                       text width=2.6cm, align=center},
        }
        \node[flow] (kstart) {Kernel starts};
        \node[flow, right=of kstart] (use) {Use ctx\\(implicit/explicit)};
        \node[flow, right=of use] (final) {shmem\_finalize()\\Destroy ctx};
        \draw[->, shorten >=2pt, shorten <=2pt] (kstart) -- (use);
        \draw[->, shorten >=2pt, shorten <=2pt] (use) -- (final);
      \end{tikzpicture}%
    }
    \caption{Device use and teardown}
  \end{subfigure}
\caption{Baseline context lifecycle for GPU-aware OpenSHMEM}
\Description{Two-panel diagram. (a) host initializes, creates an implicit context, and publishes a device-visible handle. (b) kernels use the context until finalization destroys it.}
\label{fig:context-lifecycle}
\end{figure*}
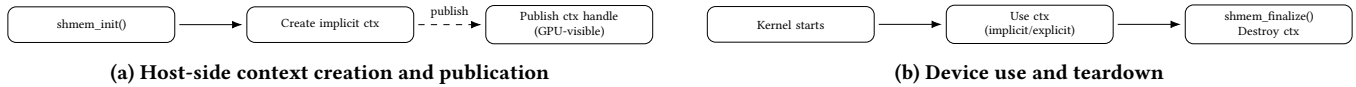
Figure~\ref{fig:context-lifecycle} captures the baseline context lifecycle from host initialization and publication through device use and teardown.
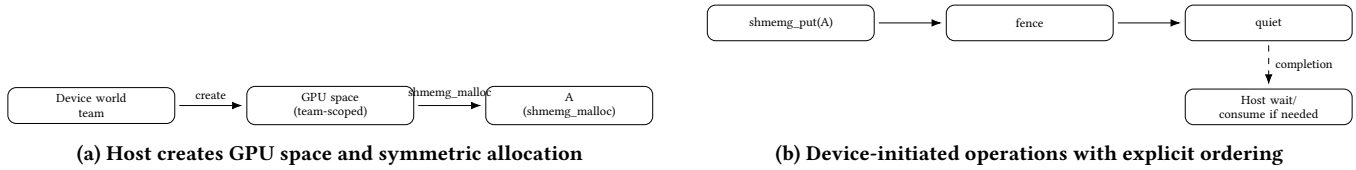
\begin{figure*}[t]
  \centering
  \begin{subfigure}[t]{0.48\textwidth}
    \centering
    \resizebox{\linewidth}{!}{%
      \begin{tikzpicture}[font=\scriptsize, node distance=12mm, >=Latex]
        \tikzset{
          flow/.style={draw, rounded corners, minimum height=6mm,
                       text width=2.6cm, align=center},
        }
        \node[flow] (team) {Device world\\team};
        \node[flow, right=of team] (space) {GPU space\\(team-scoped)};
        \node[flow, right=of space] (alloc) {A\\(shmemg\_malloc)};
        \draw[->, shorten >=2pt, shorten <=2pt]
          (team) -- node[midway, above, sloped]{create} (space);
        \draw[->, shorten >=2pt, shorten <=2pt]
          (space) -- node[midway, above, sloped]{shmemg\_malloc} (alloc);
      \end{tikzpicture}%
    }
    \caption{Host creates GPU space and symmetric allocation}
  \end{subfigure}
  \hfill
  \begin{subfigure}[t]{0.48\textwidth}
    \centering
    \resizebox{\linewidth}{!}{%
      \begin{tikzpicture}[font=\scriptsize, node distance=12mm, >=Latex]
        \tikzset{
          flow/.style={draw, rounded corners, minimum height=6mm,
                       text width=2.6cm, align=center},
        }
        \node[flow] (put) {shmemg\_put(A)};
        \node[flow, right=of put] (fence) {fence};
        \node[flow, right=of fence] (quiet) {quiet};
        \node[flow, below=8mm of quiet] (host) {Host wait/\\consume if needed};
        \draw[->, shorten >=2pt, shorten <=2pt] (put) -- (fence);
        \draw[->, shorten >=2pt, shorten <=2pt] (fence) -- (quiet);
        \draw[->, dashed, shorten >=2pt, shorten <=2pt]
          (quiet) -- node[midway, right]{completion} (host);
      \end{tikzpicture}%
    }
    \caption{Device-initiated operations with explicit ordering}
  \end{subfigure}
  \caption{Minimal usage example for the GPU space and ordering model. API names are illustrative, and exact naming is subject to auxiliary specification finalization}
  \Description{Two-panel diagram. (a) SHMEM\_DEVICE\_TEAM\_WORLD creates a team-scoped GPU space and allocates A symmetrically. (b) a kernel issues put, fence, and quiet with an optional host completion step.}
  \label{fig:space-example}
\end{figure*}
Figure~\ref{fig:space-example} illustrates the intended usage pattern: a host-created GPU space and symmetric allocation, followed by device-initiated operations with explicit ordering and host completion where required.

\subsection{Context model}
\label{sec:proposal:contexts}

\steward{The auxiliary specification separates the host default context from the device default context. The host default context exists after initialization and remains valid for host operations on \api{SHMEM\_TEAM\_WORLD}. GPU-centric operations use a device-usable default context associated with the device world team. The auxiliary specification names this context \api{SHMEM\_DEVICE\_CTX\_DEFAULT}. Explicit device contexts may be created on the host when the implementation supports them, then passed to kernels as device-visible handles.}

\steward{The discoverable property is not the existence of the host implicit context. That context is always part of the OpenSHMEM model. What must be discoverable is whether GPU-centric operations, device teams, \texttt{SHMEM\_DEVICE\_CTX\_DEFAULT}, or explicit device contexts are supported. If a context handle is device-visible, its lifetime and ordering semantics must be documented for the supported caller scope. This ensures that device-initiated operations do not violate the ordering or completion guarantees established by the OpenSHMEM model. Implementations must document valid device participation scopes and whether exposed device context handles are safe for those scopes.}

\steward{Context creation is intentionally constrained. The baseline model allows host-created contexts that may be cached in GPU-accessible memory for later use by kernels. This avoids device-side context creation during performance-critical kernels and keeps initialization collective and predictable. Optional extensions may permit device-side context creation for advanced use cases, but such features are explicitly tiered. Implementations may maintain distinct host and GPU context pools to size network connections and pipelines. Any GPU-specific context limit must be discoverable or documented.}

\subsection{RMA, atomics, synchronization, and collectives}
\label{sec:proposal:ops}

GPU-aware semantics extend existing OpenSHMEM operations to accept GPU-resident buffers when the target space is GPU-attached. \nathan{Within that umbrella and in the taxonomy described in Section~\ref{sec:comparison}, we distinguish three execution paths: host-executed operations, stream-triggered host operations, and device-initiated operations. For tier-0, host-initiated operations on GPU-resident symmetric data objects are baseline, while device-initiated and stream-triggered operations are optional. Host-executed and stream-triggered paths are both host-side invocation models. Stream-triggered variants add ordering with respect to a GPU stream or queue, but they do not change the issuing agent from host to device. An implementation may use internal GPU kernels to accelerate a host-issued or stream-triggered operation, but the observable issuing agent, ordering contract, and progress contract remain those of the host API.} The rules for symmetric allocation and pointer validity remain unchanged. Operations that require symmetric objects continue to require symmetric objects even when they reside in GPU memory.

\steward{The specification therefore requires that a GPU pointer used in an OpenSHMEM call either refer to symmetric memory in the appropriate heap or be explicitly documented as unsupported. The pointer-to-heap mapping must be unambiguous for the operation being issued. The portable model should not require device code to perform dynamic pointer classification on every operation.} \discussionMarSix{Capability discovery should present a coherent view of active GPU-aware support, device-initiated support, device-team availability, and pointer/space constraints without requiring a single fixed query API shape in this paper.}

\meetingJuneNineteen{Pointer-query routines also need an explicit access-domain contract. A host-callable team-scoped query for a GPU-attached symmetric object should not imply that the returned value is a device pointer unless the API states that property. Depending on implementation and team capability, the result may be a host-accessible alias, a device-accessible direct pointer, a pointer valid in both domains, or no pointer. Capability discovery must distinguish host accessibility, device accessibility, and direct remote-pointer availability.}

\steward{Device-initiated operations assume pre-registered symmetric heaps. Dynamic registration of arbitrary source buffers is not required. The auxiliary specification is stricter than some implementation fast paths: GPU-centric data movement operands reside in the GPU symmetric heap unless an implementation documents a capability for registered private GPU memory, immediate operands, or another extension. This is a portability rule rather than a claim that every existing implementation internally requires all local operands to be remotely addressable. The rule preserves portable addressability while leaving room for optimized implementations to expose more permissive behavior.}

\steward{Device-initiated operations are scoped to a device world team and are initially device-to-device. This device team may have the same PE membership as \api{SHMEM\_TEAM\_WORLD}, but it identifies the device-side context and atomicity domain. A device shared team denotes peers that share the relevant GPU memory locality domain and may therefore differ from \api{SHMEM\_TEAM\_SHARED}. The atomicity domain remains the team. The default device context is used unless an explicit device context is selected. Implementations must document any restriction on host-initiated operations that target device-team domains.}

Several rules are intentionally inherited unchanged from OpenSHMEM 1.6: symmetric allocation requirements, team collectivity rules, and the base meanings of \texttt{fence} and \texttt{quiet}. The new elements are GPU-attached heaps, device teams, device-visible context requirements, and explicit capability discovery for optional device-initiated operations. Contexts remain conceptually necessary for ordering and pipeline control, but the auxiliary specification does not introduce additional user-facing context management in the baseline. \discussionMarSix{A default device-usable context is the portable path, while private or specialized device contexts are treated as optional, resource-limited optimization features rather than a guaranteed one-context-per-workgroup abstraction.} Any move toward fully implicit context management is deferred to OpenSHMEM 2.0.

\steward{Atomic operations and synchronization calls inherit the same completion rules as the host model, but the specification clarifies that completion may require explicit device synchronization if the issuing thread is on the GPU. PE collectives remain collective over a team of PEs. Thread-collaborative device operations describe a separate question: which GPU threads inside one PE cooperate to issue a device operation. If a collective or RMA operation is exposed with a thread-collaborative scope, the specification mandates that all threads in the stated scope participate with identical arguments. This rule prevents divergence-related deadlocks and reflects SIMT execution constraints observed in existing GPU implementations. Scoped device RMA, workgroup or wavefront puts, and block atomics are treated as optional capabilities outside the baseline. If provided, the required participation scope and uniform-call rules must be documented.}

To reason about granularity, the specification should enumerate the scope of operations (thread, wave/warp\intel{/subgroup}, workgroup/block, grid) for RMA put/get, atomics, collectives, and ordering primitives (\texttt{fence}/\texttt{quiet}). Implementations must document which scopes they support. Atomicity does not imply ordering. Programs must use \texttt{fence}/\texttt{quiet} when ordering is required.

For blocking operations, completion is defined as local completion: the initiating thread may safely reuse the source buffer and subsequent operations issued by the same PE/context are ordered with respect to that operation. Local blocking completion does not imply that the target can be used. Remote visibility is only guaranteed after an explicit \texttt{quiet}/\texttt{barrier} and any required device or stream synchronization. For nonblocking operations, the specification preserves the existing contract that completion is established only after an explicit \texttt{quiet} or equivalent completion call. This mirrors the host model and keeps GPU-aware semantics predictable.

\begin{quote}
\small
\textbf{Normative excerpt.} GPU-resident buffers used in OpenSHMEM operations MUST be allocated from a symmetric space associated with the target team. Implementations SHOULD provide \texttt{fence}/\texttt{quiet} semantics that restore ordering guarantees on GPU memory. Implementations MAY expose device-initiated operations. Stream-ordered variants are optional follow-on capabilities discoverable at runtime.
\end{quote}

\begin{quote}
\small
\textbf{Normative excerpt.} Device collectives require uniform participation. If a device-initiated collective is provided, all threads in the stated scope MUST participate uniformly. Implementations SHOULD document restrictions on mixing host and device collectives within a team.
\end{quote}
\steward{Host and device collectives may operate on teams with the same PE membership, but they may use different default contexts and atomicity domains. Mixing host and device collectives on equivalent teams without explicit synchronization is undefined behavior.}

\subsection{Completion, ordering, and stream semantics}
\label{sec:proposal:ordering}

The auxiliary specification preserves OpenSHMEM ordering guarantees. GPU implementations may use weaker default ordering, provided \texttt{fence} and \texttt{quiet} restore ordering and completion. A \texttt{fence} orders subsequent operations with respect to prior ones in the same context, while \texttt{quiet} ensures completion of outstanding operations initiated by the caller. For device-initiated operations, this implies that consecutive gets or atomics may be unordered unless the program inserts a fence, which aligns with the behavior of current GPU implementations.

To avoid ambiguity, the caller scope of \texttt{quiet} is summarized below:
\begin{quote}
\small
\begin{tabular}{p{0.18\linewidth}p{0.72\linewidth}}
\textbf{Caller} & \textbf{\texttt{quiet} guarantees} \\
Host & Completion and remote visibility for operations initiated by the host. GPU visibility still requires explicit device synchronization (stream sync or equivalent). \\
Device & Completion for device-initiated operations in the device context. Host visibility still requires explicit host-side synchronization. \\
\end{tabular}
\end{quote}
The auxiliary specification keeps a single unscoped contract for \texttt{quiet}, \texttt{wait}, and \texttt{test}. Scoped variants remain optional future work.

\steward{Tier-0 requires implementations to define how progress is made for both host-issued and device-issued operations. That progress may be host-driven, including through a runtime progress thread. GPU-only or NIC-only progress without host involvement is optional and must be explicitly documented when provided.}

\nathan{Stream-triggered host APIs are host-issued operations whose execution is ordered relative to a GPU stream or queue. They are deferred from the initial Auxiliary 1.x release and marked as optional (e.g. See Table~\ref{tab:tiering}). The semantics below describe expected behavior if such variants are introduced in a follow-on Auxiliary 1.x revision or OpenSHMEM 2.0.} Stream-triggered variants are optional and, if provided, must define how operations are ordered within a GPU stream or queue and how they interact with other host-executed and device-initiated calls. The specification requires that stream-triggered operations not introduce deadlock with collectives or waits in a single stream. Stronger constraints (e.g., disallowing host collectives concurrent with device collectives on the same team) must be documented. This preserves flexibility while keeping the ordering contract explicit.

Implementations MUST document prohibited stream patterns and unsupported combinations.

This framing makes stream semantics explicit without over-constraining implementations. For example, a runtime may allow stream-triggered puts and gets but prohibit stream-triggered collectives, provided the restriction is documented and discoverable via capability queries.

\begin{figure}[t]
  \centering
  \fbox{\begin{minipage}{0.9\columnwidth}
  \small
  \textbf{Common pitfalls.}
  \steward{Mixing host and device collectives on the same team without explicit synchronization can deadlock. Issuing device puts in one stream and reading results in another without \texttt{fence}/\texttt{quiet} can yield stale data. Using non-symmetric GPU pointers for RMA is undefined. The auxiliary specification makes these hazards explicit and requires implementations to document any additional constraints.}
  \end{minipage}}
\caption{Ordering and stream hazards addressed by the auxiliary specification.}
\Description{A boxed callout listing pitfalls: mixed host/device collectives without sync, cross-stream access without ordering, and non-symmetric GPU pointers for RMA.}
\label{fig:pitfalls}
\end{figure}

Figure~\ref{fig:pitfalls} highlights the most common ordering and stream hazards, while Table~\ref{tab:pitfalls} maps each pitfall to the corresponding rule in the auxiliary specification.

\begin{table*}[t]
  \centering
  \caption{Common pitfalls and how the auxiliary specification addresses them.}
  \label{tab:pitfalls}
  \small
  \setlength{\tabcolsep}{4pt}
  \begin{tabular}{@{}p{0.28\textwidth}p{0.68\textwidth}@{}}
    \toprule
    Pitfall & Resolution in the auxiliary specification \\
    \midrule
    Mixed host/device collectives without sync & Requires documented restrictions and uniform participation rules to avoid deadlock and divergent execution. \\
    Cross-stream access without ordering & \texttt{fence}/\texttt{quiet} must restore ordering across streams and contexts before consumption. \\
    Host/GPU ordering assumptions & Programs must insert host stream synchronization or device fences. Ordering is explicit across domains. \\
    Host context assumed device-usable & Device operations require a supported context such as \api{SHMEM\_DEVICE\_CTX\_DEFAULT}. Otherwise device calls are undefined. \\
    Progress assumed without host involvement & Baseline does not require GPU-only progress. Optional capabilities must be discoverable and documented. \\
    Non-symmetric GPU pointers in RMA & GPU-resident operands must be allocated from symmetric spaces tied to the target team. \\
    \bottomrule
  \end{tabular}
\end{table*}

\subsection{Capability discovery}
Because GPU-aware support is not uniform across platforms, the auxiliary specification includes capability queries to determine whether GPU-resident buffers are supported for specific operations and whether device-initiated calls are available. These queries allow applications to select safe code paths and provide graceful fallbacks. Capabilities are defined per team and may differ across teams. Applications should re-query after creating a new team and may cache results for the lifetime of a team. Capability queries MUST distinguish GPU-aware host-initiated support from GPU-initiated device APIs, and SHOULD report optional stream/queue ordering variants separately.

Capability queries also expose optional tiers. Programs can discover whether a team supports device-initiated collectives, explicit device contexts, or later stream-triggered host interfaces. Implementations MAY also expose a locality query for the device shared team that reports the subset of PEs in the relevant GPU locality domain. This set may differ from \texttt{SHMEM\_TEAM\_SHARED} on non-fully-connected topologies and is intended as an optimization aid, not a semantic requirement. Implementations MAY provide implementation-defined controls to enable or disable GPU-aware and GPU-initiated features. Defaults are implementation-defined, but capability queries must reflect the active mode. This keeps the specification flexible while allowing applications to use advanced features when present.
Implementations MAY also expose a capability that reports the active GPU-initiated communication path (e.g., CPU proxy, direct intra-node link, or NIC offload). This is an optimization hint and does not change the baseline semantics.

\subsection{Error Model}
An error model is outside the scope of this auxiliary specification and is deferred to OpenSHMEM 2.0. We will adhere to the following OpenSHMEM conventions: undefined behavior is permitted if the application violates symmetric allocation rules, context lifetime, or collective participation requirements. Implementations may provide additional diagnostics, but the baseline semantics prioritize portability and predictable behavior.

\section{Decision Points and Tiering}
\label{sec:decisions}

The auxiliary specification is designed around explicit decision points that balance portability with performance. We organize these decisions as a staged roadmap so that the baseline can be standardized quickly while later tiers absorb device-centric and hardware-specific features without overloading the first release.

Key decision points include:
\begin{itemize}
  \item \textbf{Context visibility and granularity}: Whether implicit contexts are device-visible by default, and whether explicit contexts can be specialized per workgroup.
  \item \textbf{Adoption path framing}: Whether the auxiliary specification is CPU-first with optional GPU features or dual-track with CPU and GPU on equal footing.
  \item \textbf{Initialization and feature selection}: Whether requested feature subsets are selected at initialization time and how those controls are exposed. The paper leaves the exact API form open. Post-init capability queries are the portable contract.
  \item \textbf{Memory spaces}: Whether a two-space model (CPU and GPU) is sufficient, how registration keys are managed across spaces, and how the default heap is selected and reported. Heap granularity and coherency traits remain tuning properties.
  \item \textbf{Space lifecycle and heap selection}: Auxiliary 1.x assumes static heaps without dynamic space creation and a two-heap model (CPU and GPU). The default heap type (CPU vs GPU) must be discoverable, with an explicit allocation path to select the non-default heap.
  \item \steward{\textbf{GPU teams and locality}: The device world team scopes device-initiated operations. The device shared team denotes peers in the relevant GPU locality domain and may impose stricter atomicity restrictions than host-visible shared teams.}
  \item \textbf{Device mapping and transport path queries}: Whether optional queries should report GPU-space to device association (when a PE controls multiple GPUs), GPU-team availability, and the active GPU-initiated communication path (proxy vs direct vs NIC offload).
  \item \meetingJuneNineteen{\textbf{Pointer accessibility}: Whether team pointer queries return a host pointer, device pointer, both, or none. The baseline should not imply that a pointer valid in one access domain is valid in another unless that property is explicitly reported.}
  \item \textbf{Context exposure vs implicit contexts}: Contexts remain necessary for ordering and pipeline control, but OpenSHMEM 2.0 could make them implicit. The auxiliary specification specifies context to follow 1.x semantics.
  \item \textbf{Context resources and specialization}: A default device-usable context is the portable path. Private or specialized device contexts are optional, resource-limited optimization features and should not be treated as a guaranteed one-context-per-workgroup abstraction.
  \item \textbf{Ordering and completion}: Guarantees of \texttt{fence}/\texttt{quiet} on GPU memory, interaction with stream-triggered host operations (if introduced), and the boundary between unscoped baseline semantics and future scoped variants.
  \item \textbf{Progress model}: Whether forward progress requires host involvement or can be guaranteed by GPU or NIC offload.
  \item \steward{\textbf{Device initiation}: Whether GPU threads can issue RMA, atomics, and collectives, under what constraints, and how resource-limited features are discovered. For kernel-initiated put-signal, signal and target are expected in the GPU symmetric heap. Documented extensions may allow immediate or private source operands.}
\end{itemize}

To reduce ambiguity for implementers, we make the following clarifications about baseline intent and optional capabilities:
\begin{itemize}
  \item \steward{\textbf{Baseline}: Host-executed operations on GPU-resident symmetric buffers are required. Device-initiated operations are optional and discoverable.}
  \item \textbf{Device initiation does not imply GPU-initiated networking}: Device-initiated operations may be implemented via host proxy mechanisms or direct GPU transport. The observable semantics must match either way.
  \item \textbf{Context resources are not prescribed}: \ac{Shared}-style (e.g., \ac{IPC}) or \ac{Dedicated}-style (e.g., \ac{GDA}) mappings are both valid if ordering and completion semantics are honored.
  \item \steward{\textbf{Context visibility is explicit}: An implicit context must exist for host use. GPU-centric operations require a supported device context. Explicit device contexts are optional and must be safe at the documented granularity when exposed.}
  \item \textbf{Progress is allowed to be host-driven}: GPU- or NIC-driven progress is optional. 
  \item \textbf{Eliminating the need for user context management complexity is an OpenSHMEM 2.0 design question.}
  \item \textbf{Stream APIs are a follow-on}: Stream-triggered host variants are optional and deferred from the initial Auxiliary 1.x release. No stream ordering is implied beyond \texttt{fence}/\texttt{quiet} unless explicitly documented. The working plan is to address stream APIs in a later tier or OpenSHMEM 2.0.
\end{itemize}

We also make the implementer clarifications explicit to avoid confusion:
\begin{itemize}
  \item \textbf{No GPU-initiated networking requirement}: Tier-0 conformance MUST NOT require GPU-resident transport endpoints or GPU-initiated NIC progress.
  \item \steward{\textbf{No mandatory GPU-only progress}: Host-driven progress is permitted in Tier-0 for both host-issued and device-issued operations. GPU-only progress is an optional capability.}
  \item \textbf{Dual-track adoption is allowed}: Implementations may present GPU-aware features as optional add-ons to a CPU-first stack or as primary functionality in a GPU-first stack, provided capabilities are accurately reported.
  \item \textbf{No progress guarantee for persistent kernels}: The auxiliary specification does not define progress guarantees for persistent GPU kernels without host participation.
  \item \textbf{No standardization of resource mappings}: The auxiliary specification does not prescribe how contexts map to queue pairs, endpoints, or NIC resources.
  \item \textbf{No ABI changes}: The auxiliary specification does not require ABI changes to OpenSHMEM 1.6, thereby preserving backward compatibility.
  \item \textbf{No required stream semantics}: Stream-ordered variants are optional and must be explicitly documented when provided.
\end{itemize}

Tier-0 requires GPU-aware memory spaces, host-initiated operations on GPU-resident buffers, and well-defined ordering and completion semantics. Tier-1 adds device-centric features that can be exposed today without changing the baseline portability contract. Tier-2 captures more specialized follow-on features that need additional implementation maturity or broader consensus. Additional tiers may emerge later as the community gains experience.

Table~\ref{tab:tiering} summarizes the staged tiers used throughout the specification.

\begin{table*}[t]
  \centering
  \caption{Tiering summary for auxiliary specification features}
  \label{tab:tiering}
  \footnotesize
  \setlength{\tabcolsep}{4pt}
  \begin{tabular}{p{0.30\textwidth}p{0.30\textwidth}p{0.30\textwidth}}
    \toprule
    Tier-0 (portable baseline) & Tier-1 (device-centric optional) & Tier-2 (specialized/future optional) \\
    \midrule
    GPU-aware buffers & Device-initiated operations & Stream-triggered host APIs \\
    Two-space memory model & Default device-usable context & Scoped wait/test/quiet variants \\
    Discoverable default heap & Device-team locality queries & Per-workgroup/private context specialization \\
    \texttt{SHMEM\_DEVICE\_TEAM\_*} semantics & Explicit device contexts (if exposed) & Richer device-team atomicity specialization \\
    Host default and device default contexts & Host/device path reporting & GPU-driven progress or stronger offload assumptions \\
    fence/quiet semantics & Host-driven or proxy-backed progress & Additional spaces or dynamic resource models \\
    Capability discovery & Device collectives (if exposed) & Future vendor-specific extensions \\
    \bottomrule
  \end{tabular}
\end{table*}

\section{Conformance and Implementation Mapping}
\label{sec:conformance}

Conformance is central to making the auxiliary specification actionable. Each MUST/SHOULD requirement in the proposed text maps to at least one test that distinguishes correct behavior from undefined or implementation-specific behavior. These tests serve both as validation for implementers and as portable checks for applications.

Representative conformance tests include:
\begin{itemize}
  \item \textbf{Context visibility}: \steward{Verify that a default device context is available when GPU-centric support is reported, and that explicit contexts can be used safely from device code when exposed.}
  \item \textbf{Space symmetry}: Allocate a GPU space collectively and confirm that symmetric objects are addressable and valid across PEs.
  \item \textbf{GPU space at init}: If a GPU-first implementation provides a default GPU symmetric heap at initialization, verify capability reporting and that allocations are symmetric without explicit space creation.
  \item \textbf{Ordering/completion}: Confirm that \texttt{quiet} implies completion for GPU-resident operations and that \texttt{fence} enforces ordering within a context.
  \item \textbf{Stream ordering}: If stream-triggered APIs are provided, verify ordering within a stream and document constraints for host/device collective interactions.
  \item \meetingJuneNineteen{\textbf{Pointer accessibility}: If direct pointer queries are exposed, verify that the returned pointer is valid only in the documented host, device, or shared access domain.}
  \item \textbf{Progress}: Distinguish implementations that require host progress from those that provide GPU/NIC-driven progress.
\end{itemize}

We envision these tests living alongside existing OpenSHMEM conformance suites, with an auxiliary repository for rapid iteration during the specification phase and migration into the main test suite as the auxiliary spec stabilizes. This structure keeps the barrier to adoption low while ensuring the tests become a shared community asset.

Existing implementations can be mapped to these tests as follows: NVSHMEM and rocSHMEM provide device-initiated operations with explicit ordering primitives, while \intel{\ishmem{}} focuses on GPU-centric semantics with different device visibility and context handling. This mapping provides a concrete way to validate both baseline and optional features without assuming a single vendor-specific behavior.

In practice, this means a Tier-0 implementation should pass tests for symmetric GPU allocation, fence/quiet ordering, and host-initiated RMA on GPU buffers. Later tiers add opt-in conformance extensions for device-side contexts, stream-triggered host operations, and GPU-driven progress. The mapping also highlights implementation differences. NVSHMEM exposes stream-triggered variants and device-side collectives. rocSHMEM specifies that \texttt{\_wg} and \texttt{\_wave} (e.g. \api{rocshmem\_barrier\_all\_wave}) routines require all threads in the workgroup or wavefront to call with the same parameters, while NVSHMEM provides block-scoped collectives via \api{nvshmemx\_*\_block}. \intel{\ishmem{} provides \api{\_work\_group} collectives that accept a \api{sycl::group} argument covering both workgroup and subgroup scopes~\cite{brooks2024intel}. We treat all three as optional workgroup-scoped collectives in conformance, with participation rules verified per implementation~\cite{rocshmem,nvshmem,brooks2024intel}.} These behaviors can be captured as optional conformance extensions rather than baseline requirements.

\section{Impact and Practical Evaluation}
\label{sec:evaluation}

This paper focuses on specification design rather than benchmarking, but the proposed auxiliary specification has direct practical impact. By defining a portable GPU-aware baseline, applications can avoid vendor-specific branches for symmetric allocation and RMA semantics while retaining optional paths for GPU-initiated operations. This reduces maintenance burden and improves how performance tools and correctness checkers reason about GPU-resident communication. \reviewers{It also gives implementers a common vocabulary and conformance target for portable SHMEM behavior across deployed and emerging implementations.} Based on the implementation survey, most Tier-0 behaviors (symmetric GPU allocation, fence/quiet ordering, GPU-aware buffers) already exist in current vendor-specific implementations, so conformance should require limited changes beyond capability reporting and clarifying device visibility rules. The staged tiering also supports incremental validation: Tier-0 tests establish portability, while later tiers add opt-in checks for device-centric and specialization features.

\section{Discussion and Roadmap}
\label{sec:discussion}

The auxiliary specification deliberately prioritizes portability while allowing advanced implementations to expose GPU-initiated communication. This tension between portability and performance is unavoidable in heterogeneous systems: strict guarantees can limit performance on weakly ordered devices, while aggressive offload can introduce semantics that are hard to standardize.

The roadmap is therefore staged. Tier-0 provides GPU-aware semantics that can be implemented broadly across vendors and transports. Tier-1 adds device-centric capabilities such as device-initiated operations, default device-usable contexts, and locality-aware device teams without changing the core portability contract. Tier-2 captures more specialized features, including stream-triggered host APIs, scoped synchronization variants, and deeper context/resource specialization. \discussionMarSix{The tier structure is intentionally open-ended: additional tiers may emerge as hardware support and implementation practice converge.} Heap granularity and coherency knobs remain implementation-specific in Auxiliary 1.x, while more specialized device-side semantics are deferred to later tiers or OpenSHMEM 2.0.

A key open question is how to express progress guarantees for persistent kernels and long-lived GPU workloads. The auxiliary specification intentionally does not mandate GPU-only progress, but it requires that implementations document when host involvement is necessary. This keeps the model honest while enabling innovation in GPU-driven progress.
Another open question is whether contexts should remain explicit or become implicit in a future OpenSHMEM 2.0 model. The auxiliary specification preserves existing context semantics to avoid breaking 1.x, but it records the goal of reducing user-facing context management as a 2.0 design target. Other likely OpenSHMEM 2.0 topics include stronger locality-aware team semantics, richer scoped synchronization, and any standardization of stream-triggered host interfaces that the community decides to pursue.
Naming conflicts appear minimal for the auxiliary APIs (primarily \texttt{shmem\_malloc}). If conflicts arise, a shim/rename layer remains a viable migration bridge. The paper does not treat the current auxiliary proposal as an API lock-in document: its role is to define the portable core, document the decision points, and show how future tiers can evolve from a stable baseline.

\section{Conclusion}
\label{sec:conclusion}

We propose a GPU-OpenSHMEM auxiliary specification that provides a vendor-neutral, backward-compatible path for GPU-aware communication. By grounding the specification in evidence from existing implementations, we define a minimal memory spaces model, a context framework that supports device visibility, and explicit ordering and completion rules suitable for GPU memory. The decision-point and tiering structure clarifies what is required for portability and what is optional for advanced hardware and runtimes. This auxiliary specification is intended as a practical step toward a unified OpenSHMEM 2.0. The combination of a clear baseline, optional GPU-centric features, and a conformance-oriented test plan provides a concrete way forward for both vendors and application developers.

\begin{acks}
This work was funded through Strategic Partnership Projects Funding Office via Los Alamos National Laboratory with IAN 619215901 on the project ``OpenSHMEM - Standardized API for parallel programming in the Partitioned Global Address Space.'' This manuscript has been authored by UT-Battelle, LLC under Contract No. DE-AC05-00OR22725 with the U.S. Department of Energy.

Generative AI tools were used to assist with editorial suggestions, phrasing, and grammar. 
\end{acks}

\bibliographystyle{ACM-Reference-Format}
\bibliography{software}

\appendix

\end{document}